\documentclass[11pt,preprint]{aastex}

\usepackage{color}
\usepackage{natbib}
\usepackage{amsmath}
\tightenlines
\slugcomment{}

\def \bv         {{\bf v}}

\newcommand \beq        {\begin{equation}}
\newcommand \beqa	{\begin{eqnarray}}

\newcommand \cm         {\,{\rm cm}}

\newcommand \eeq	{\end{equation}}
\newcommand \eeqa	{\end{eqnarray}}

\newcommand \erg	{\,{\rm erg}}

\newcommand \eV 	{\,{\rm eV}}

\newcommand \gtsim	{\gtrsim}		 
\newcommand \Ha 	{{\rm H}}
\newcommand \He	        {{\rm He}}

\newcommand \IH         {I_{\rm H}}

\newcommand \kB         {k_{\rm B}}

\newcommand \K  	{\,{\rm K}}

\newcommand \ltsim	{\lesssim}		 

\newcommand \nH         {n_{\rm H}}

\newcommand \s	        {\,{\rm s}}

\newcommand \xtimes     {{\!\,\times\!\,}}

\newcommand{\oldtext}[1]{}
\newcommand{\newtext}[1]{{{#1}}}

\countdef\decade=200
\advance\decade by \year
\countdef\hours=201
\advance\hours by \time
\divide\hours by 60
\countdef\mins=202
\advance\mins by \hours
\multiply\mins by 60
\multiply\hours by 100
\countdef\miltime=203
\advance\miltime by \hours
\advance\miltime by \time
\advance\miltime by -\mins

\begin{document}

\title{%
        \vspace*{-3.0em}
        {\normalsize\rm {\it The Astrophysical Journal}, {\bf 862}:30 (2018 July 20)}\\ 
        \vspace*{1.0em}
        Electron Energy Distributions in \ion{H}{2} Regions
        \newtext{and Planetary Nebulae}:
        $\kappa$-Distributions Do Not Apply
	}

\author{B.~T. Draine and C.~D. Kreisch}\affil{Princeton University Observatory,
        Peyton Hall, Princeton, NJ 08544-1001, USA;
        draine@astro.princeton.edu, ckreisch@astro.princeton.edu}

\begin{abstract}
Some authors have proposed that electron energy
distributions in \ion{H}{2} regions 
\newtext{and planetary nebulae}
may be significantly nonthermal,
and $\kappa$-distributions have been suggested as being appropriate.
Here it is demonstrated that
the electron energy distribution function 
is extremely close to a Maxwellian up to electron kinetic energies
$\sim 13\eV$ \newtext{in \ion{H}{2} regions, and up to $\sim16\eV$
in planetary nebulae}:
$\kappa$-distributions are inappropriate.
The small
departures from a Maxwellian have negligible effects on line ratios.
When observed line ratios in \ion{H}{2} regions deviate from
models with a single electron temperature, it must arise from
spatial variations in electron temperature, rather
than local deviations from a Maxwellian. 
\end{abstract}

\section{Introduction}

Observed ratios of collisionally-excited emission lines
from \ion{H}{2} regions \newtext{and planetary nebulae}
in some cases depart from the predictions
for a plasma with thermally-distributed electrons with a single
electron temperature $T_e$.
Discrepancies are also found when comparing $T_e$ determined
from ratios of collisionally-excited lines 
(e.g., [\ion{O}{3}]4364/[\ion{O}{3}]5008) with $T_e$ estimated
from the Balmer or Paschen discontinuities in hydrogen recombination
spectra.
Finally, abundances estimated from the faint recombination
lines of heavy elements often differ substantially from abundances
estimated from collisionally-excited lines.

\citet{Peimbert_1967} proposed that the observed line ratios
\newtext{in \ion{H}{2} regions}
are due to  spatial variations in $T_e$ within the 
\newtext{photoionized gas} -- 
which Peimbert referred to as ``temperature fluctuations'' --
and line ratios are sometimes
interpreted within this framework
\citep[e.g.,][]
{Louise+Monnet_1969,
Liu+Danziger_1993,
Kingdon+Ferland_1995,
Gruenwald+Viegas_1995,
Perez_1997,
Kingdon+Ferland_1998,
Binette+Luridiana_2000,
ODell+Peimbert+Peimbert_2003,
Zhang+Ercolano+Liu_2007,
Oliveira+Copetti+Krabbe_2008}.
\citet{Esteban_2002} reviewed arguments for and against the existence
of such temperature variations within ionized nebulae.
\citet{Binette+Matadamas+Hagele+etal_2012} claimed that observed
line ratios in \ion{H}{2} regions could {\it not} be accounted for by
temperature inhomogeneities, and argued that the observed line ratios
required either unusual heating mechanisms (e.g., shock waves),
metallicity inhomogeneities, or electron energy distributions given
by the $\kappa$-distribution instead of the Maxwell-Boltzmann distribution.

Some authors have proposed that the energy
distribution functions may {\it locally}
depart from a Maxwellian distribution.
\citet{Giammanco+Beckman_2005} proposed that ionization by cosmic
rays could be responsible for non-Maxwellian electron energy distributions.
\citet{Nicholls+Dopita+Sutherland_2012} have argued that the electron
energies obey a $\kappa$-distribution, and
\citet{Nicholls+Dopita+Sutherland+etal_2013},
\citet{Dopita+Sutherland+Nicholls+etal_2013},
and
\citet{Mendoza+Bautista_2014}
have calculated emission line
ratios assuming the electron energies to be $\kappa$-distributed.
\citet{Zhang+Liu+Zhang_2014} discussed $\kappa$-distributed electrons
in planetary nebulae.

The $\kappa$-distribution has been found to describe the distribution of
electron energies in interplanetary plasmas
\citep{Pierrard+Lazar_2010}, where the low densities
of the solar wind are insufficient to establish a thermal distribution given the
high temperatures and short time scales associated with flow from
the Sun to the Earth's orbit.

By contrast, it had generally been assumed that for
conditions in \ion{H}{2}
regions, elastic scattering is fast enough that
the electron energy distribution should be very close to
a Maxwell-Boltzmann distribution
\citep{Spitzer_1941,Spitzer_1968}.
\citet{Ferland+Henney+ODell+Peimbert_2016} estimated the time scales
and distances on which suprathermal electrons in an \ion{H}{2}
region would be thermalized, and argued that the speed of thermalization
suggested that it was unlikely that such electrons could affect
the collisionally-excited forbidden lines, but noted that there
did not appear to be any numerical calculations examining this in detail.

Because of the recent proposals that electron energy distributions
in \ion{H}{2} regions may be significantly non-Maxwellian,
we revisit this problem here.
From first principles, 
we explicitly solve for the steady-state distribution of electron energies
in a partially-ionized 
plasma where heating of the plasma is done by
injection of energetic photoelectrons
resulting from photoionization of H atoms that
have been formed by radiative recombination,
and where electron cooling is accomplished by various processes, including
radiative recombination, free-free emission, and collisional
excitation of species such as \ion{O}{3} and \ion{S}{3}.

We show that the electron energy distribution relaxes rapidly
to a steady-state distribution that is very close to a Maxwellian.  
Slight departures from a Maxwellian do occur, but are
far too small to noticeably affect observed emission line ratios.

The paper is organized as follows.
The statistical problem for the steady-state case
is formulated in Section \ref{sec:formulation}.
The atomic processes are reviewed in Section \ref{sec:physical processes}:
the Coulomb scattering responsible for thermalization, 
radiative recombination of hydrogen followed by injection of photoelectrons,
and cooling by free-free emission and collisional excitation of
ions.

In Section \ref{sec:steady-state} we present the
steady-state solution for the electron
energy distribution in an \ion{H}{2} region with a representative
spectrum of photoelectrons.
The steady-state solutions are shown to be extremely close to
a Maxwell-Boltzmann distribution, with only small departures at very
suprathermal energies.  The $\kappa$-distribution does {\it not}
describe the electron energy distribution.

In Section \ref{sec:relaxation}
we study the time evolution of an electron
distribution starting from an initial distribution that is far from
thermal; relaxation to a near-thermal distribution takes place extremely
rapidly.
For conditions in the Orion Nebula, thermal relaxation
is accomplished in $\sim$$30\sec$.
\newtext{In Section \ref{sec:PN} we model the electron energy distribution in
a prototypical planetary nebulae, photoionized by radiation from
a $10^5\K$ star.} 

We conclude that electron energy distributions in \ion{H}{2}
regions and planetary nebulae will be locally very close to
Maxwell-Boltzmann, \newtext{except for a high energy tail containing only
a very small fraction of the electrons.}  
There is no basis for using the $\kappa$-distribution
to describe the distribution of electron energies in \ion{H}{2} regions
\newtext{and planetary nebulae}. 

Certain technical details concerning the treatment of elastic
scattering are presented in the Appendix.

\section{\label{sec:formulation}
         Formulation of the Problem}

To study the distribution of electron energies, we define
$N$ electron energy ``bins'' $j=1,...,N$, where
bin $j$ includes kinetic energies $(j-1)\delta E < E < j\delta E$,
where $\delta E \equiv E_{\rm max}/N$.

Let $n_e$ be the electron density, and let $P_j$ be the
fraction of the electrons having energies
in bin $j$.
Let $n_X$ be the number density of other species $X$ present.
An electron in bin $j$ can be scattered to a bin $k\neq j$
by elastic scattering off electrons or ions, or by inelastic
scattering by atoms or ions, where some of the initial energy $E_j$
goes into excitation of the atom or ion.
Scattering of the electron by ions can result in energy loss by
bremsstrahlung.
Radiative recombination can remove the electron from bin $j$, and
photoionization will inject new electrons into the energy distribution.

Transition rates for a number of distinct processes are required:
\begin{itemize}
\item $A_{kij}P_j$ = probability per unit time that an electron with
energy $E_i$ will gain energy $E_k$ from an electron with
energy $E_j$, becoming an electron with
energy $E_i+E_k$.
The electron with
energy $E_j$ becomes an electron with energy $E_j-E_k$.
\item $B_{ji}$ = probability per unit time that an electron
with energy $E_i$ will radiate a continuum photon $h\nu=E_i-E_j$
as a result
of free-free scattering by an ion $X^+$:
\beq
X^+ + e^-(E_i) \rightarrow X^+ + e^-(E_j) + h\nu
~.
\eeq
\item $C_{ji}$ = probability per unit time of inelastic
scattering of an electron of energy $E_i$
by one of the species in the gas (e.g., \ion{S}{2}, \ion{N}{2},
\ion{O}{3}) with energy loss $E_i-E_j$ corresponding to excitation
of the target species.  The transition matrix $C_{ji}$ should include
all important inelastic excitation channels.
\item $R_i$ = probability per unit time that an electron with
energy $E_i$ will undergo radiative recombination
\beq
X^+ + e^-(E_i) \rightarrow X + h\nu
~.
\eeq
\end{itemize}
The only important species to consider for $X$ are H and He.
We assume that radiative recombination is balanced by photoionization
\beq
X + h\nu \rightarrow X^+ + e^-(E)
~,
\eeq
where $E=h\nu-I_X$.  Let $\phi_j$ be the probability that the
photoionization event has $E \in [E_j-\delta E/2,E_j+\delta E/2]$.
The assumption that recombination is instantaneously balanced by
photoionization is not exact for time-dependent calculations, but remains
a good approximation if the density of ionizing photons is high so
that the time for photoionization is short, and the neutral fraction
is very small (i.e., high values of the ``ionization parameter'').

We assume that $E_{\rm max}$ is large enough that
transitions to and from electron energies $E>E_{\rm max}$ can
be neglected, and $\sum_{j=1}^N P_j=1$.
$E_{\rm max}$ should be large enough that photoionization events producing
photoelectrons 
with $E > E_{\rm max}$ can be neglected.
This corresponds to assuming that the ionizing spectrum does
not extend beyond $h\nu=\IH+E_{\rm max}$.
\newtext{
For photoionization by massive stars, 
we will typically take $E_{\rm max}=25\eV$.
This value of $E_{\rm max}$ allows for injection of photoelectrons
resulting from photoionization by photons up to $h\nu=\IH+E_{\rm max}=38.6\eV$,
which includes over 99\% of the ionizing radiation from a star
with $T_\star=35000\K$ (spectral type O8V).
To study the electron energy distribution for conditions in 
planetary nebulae, we set
$E_{\rm max}=75\eV$, allowing for photoionization of H by photons up to
$h\nu=88.6\eV$, which includes over 99\% of the ionizing radiation for
a $T_\star=10^5\K$ blackbody spectrum.}

For every $i$, we have
\beqa \nonumber 
\frac{dP_i}{dt} &=&
\left(\sum_{j=1}^N R_j P_j\right) \phi_i
-
R_i P_i
+
\sum_{j=i+1}^N \left(B_{ij}+C_{ij}\right)P_j
-
\sum_{j=1}^{i-1}\left(B_{ji}+C_{ji}\right)P_i
\\ \nonumber
&&+\sum_{j=1}^{i-1} ~ \sum_{k=i-j+1}^N   A_{i-j,j,k} P_j P_k 
+\sum_{k=i+1}^N \sum_{j=1}^{N-(k-i)} ~ A_{k-i,j,k} P_j P_k 
\\ \label{eq:dPdt}
&&-\sum_{j=1}^{N-i} ~ \sum_{k=j+1}^N A_{j,i,k} P_i P_k 
-\sum_{j=1}^{i-1} ~ \sum_{k=1}^{N-j} A_{j,k,i} P_k P_i 
~.
\eeqa

The physics is contained in
$A_{ijk}$, $B_{ij}$, $C_{ij}$, $R_i$, and $\phi_i$.

\section{\label{sec:physical processes}
         Physical Processes}
\subsection{Elastic Scattering}

Elastic scattering is the fastest and most important process, and
accurate calculation of the $O(N^3)$ nonzero elements of $A_{ijk}$
is challenging.
For the Coulomb interaction, large impact parameter weak
scattering events make the dominant contribution to energy transfer.
These weak-scattering events must be treated with some care.
The key length scale is the plasma Debye length
\beq
L_{\rm Debye} = 690 \left(\frac{T_e}{10^4\K}\right)^{1/2}
\left(\frac{n_e}{\cm^{-3}}\right)^{-1/2}\cm
~~.
\eeq
On scales short compared to $L_{\rm Debye}$ the electrons and ions
are located randomly, but on longer length scales electrons and ions
are correlated, leading to screening.

In the center-of-mass frame, the electron-electron scattering problem
is entirely characterized by the center-of-mass energy and
the impact parameter $b$.
For impact parameter $b<L_{\rm Debye}$ we assume that screening effects
can be neglected and the interaction
is described by
classical 2-body Rutherford scattering, with
differential scattering cross section \citep[e.g.][]{Landau+Lifshitz_1976}
\beq
\frac{d\sigma}{d\Omega} = \left(\frac{e^2}{4E_{\rm CM}}\right)^2
\frac{1}{\sin^4(\theta/2)}
~,
\eeq
where $\theta$ is the scattering angle in the center-of-mass frame, and
$E_{\rm CM}$ is the center-of-mass energy.
For $b>L_{\rm Debye}$, we assume that screening is highly effective,
and these events will be neglected entirely.
Our approximations will overestimate the contribution of scattering
events with impact parameters $0.5L_{\rm Debye} \ltsim b < L_{\rm Debye}$,
while underestimating (by entirely neglecting) the contribution of scattering
events with $b>L_{\rm Debye}$.

Consider scattering between electrons with energy $E_j$ and $E_k$.
For $i>1$ we take $A_{ijk}$ to be defined so that
$A_{ijk}E_i$ is the correct rate of energy transfer in scattering
events where the electron with initial energy $E_j$ scatters into
states with final energy in 
$[E_{i+j}-\frac{\delta E}{2},E_{i+j}+\frac{\delta E}{2}]$.
For chosen $E_j$ and $E_k$, computation of $A_{ijk}$ requires
determining the rate of collisions 
such that the electron with initial energy $E_j$ emerges
with kinetic energy increased by 
$\Delta E \in [i-\frac{1}{2},i+\frac{1}{2}]\delta E$.

The smallest energy transfer events must be treated carefully, because they
dominate the total loss rate.  For $A_{1jk}$ and $k>j$ we include
all events where electron $k$ loses energy with 
$|\Delta E|<\frac{3}{2}\delta E$.

We explicitly calculate scattering rates $A_{ijk}$ where $E_k>E_j$ and
particle $j$ gains energy from particle $k$.  For $j\leq k$ we obtain
the $A_{ijk}$ using detailed balance
(see Appendix \ref{app:detailed balance}).

\begin{figure}[ht]
\begin{center}
\includegraphics[angle=0,width=10.0cm,
                 clip=true,trim=0.5cm 5.0cm 0.5cm 2.5cm]
{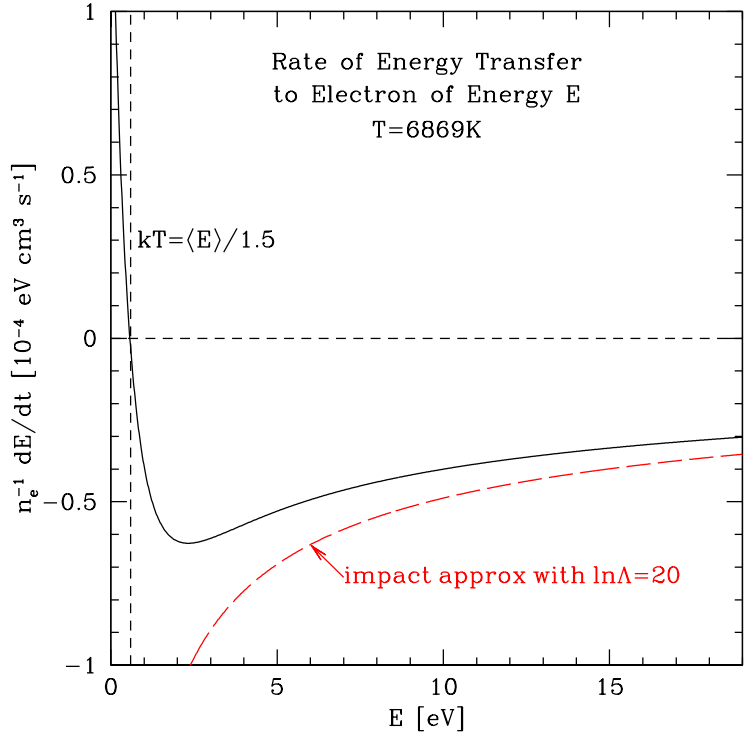}
\caption{\label{fig:dE/dt}\footnotesize
         Average rate of energy gain or loss for an electron of
         energy $E$ due to elastic scattering in a thermal plasma of
         temperature $T=6869\K$, the temperature corresponding to
         thermal equilibrium for the conditions considered below.
         }
\end{center}
\vspace*{-0.5cm}
\end{figure}

Further details of our treatment of elastic scattering are provided
in Appendices \ref{app:ee scattering}--\ref{app:detailed balance}.  
Figure \ref{fig:dE/dt} shows the net rate of
energy change $dE/dt$ due to elastic scattering only,
\beq
\left(\frac{dE}{dt}\right)_i 
= \sum_{j=1}^N \sum_{k=1}^N \left(A_{kij}-A_{kji}\right)P_j E_k
\eeq
as a function of electron energy $E_i$, for
a thermal distribution $P_j$.
As expected, $dE/dt$ is zero for $E=\langle E\rangle = 1.5 \kB T$,
where $\kB$ is Boltzmann's constant; lower energy
electrons systematically gain energy, while higher energy electrons
on the average lose energy to their neighbors.
The broken line in Figure 1 shows the rate of energy loss for an electron
moving through a zero temperature plasma, calculated
using the impact approximation
\citep[e.g., \S2.2.1 of][]{Draine_2011a}, 
which should be valid for energies
$E\gg \langle E \rangle$:
\beq \label{eq:impact approx}
\frac{dE}{dt} = - \frac{4\pi n_e e^4}{(2 m_e E)^{1/2}}\ln\Lambda
~,
\eeq
where $\ln\Lambda \approx 20$ is the usual Coulomb logarithm.
For $E \gtsim 10 \eV$ our calculated rate of net energy change is
in fairly good agreement with Eq.\ (\ref{eq:impact approx}).

\subsection{\label{sec:recombination}
            Radiative Recombination and Injection of Photoelectrons}

We assume ``case B'' recombination, neglecting radiative recombination
directly to the ground state.
If the case B
radiative recombination rate coefficient in a thermal plasma
is approximated by a power-law
$\alpha_B \approx 2.59\times10^{-13}T_4^{-0.81}\cm^3\s^{-1}$
\citep[see, e.g.,][]{Draine_2011a},
then it follows that the
rate coefficient for recombination of an electron of energy $E$ varies as
\beq
\alpha(E) = 1.55\times10^{-13}\left(\frac{E}{\eV}\right)^{-1.31}\cm^3\s^{-1}
~.
\eeq
Thus,
\beq
R_i = 1.55\times 10^{-13}\left(\frac{E_i}{\eV}\right)^{-1.31}\cm^3\s^{-1}
\times n(\Ha^+)
~.
\eeq
Recombination preferentially removes low-energy electrons.

In the steady-state, every recombination is balanced by a photoionization.
The energy distribution $\phi_i$ of the photoelectrons is determined by
the spectrum of the absorbed photons.  This will vary with distance
from the exciting star, with softer photons being preferentially
absorbed near the star, hardening the spectrum with increasing
distance from the star.
Near the star the photoionization rate is dominated by the
stellar photons with energies just above $\IH$.
The photon spectrum hardens as one moves away from the star, and
near the outer edge of the \ion{H}{2} region the ionizing spectrum
is hardest, and the mean photoelectron
energy is highest.
If every ionizing photon is absorbed somewhere, then the overall average
spectrum will be that of the ionizing photons emitted by the star.
For simplicity, we approximate the stellar spectrum by a
blackbody of temperature $T_\star$, and assume all photoionizations
come from hydrogen.
Therefore, for stellar temperature $T_\star$, 
we approximate the spectrum of photoelectrons by
\newtext{the average spectrum}
\newtext{
\beq \label{eq:phi_i}
\phi_i = 
\frac{(\IH+E_i)^2/ \left[e^{(\IH+E_i)/\kB T_\star}-1\right]}
     {\sum_{j=1}^N (\IH+E_j)^2/\left[e^{(\IH+E_j)/\kB T_\star}-1\right]}
~.
\eeq

}

If the exciting star is hot enough to appreciably ionize He, then
helium ionization and recombination will also occur.  On the one hand,
this provides an additional heating mechanism; on the other hand,
because of the higher ionization threshold, the photoelectrons
will be less energetic.
For simplicity, we will ignore both photoionization and recombination
of helium.
With $n(\He^+)/n(\Ha^+)\approx 0.1$,
we are thereby underestimating the overall photoelectric heating rate
by perhaps $\sim$10\% for a plasma where the helium is predominantly
\ion{He}{2}.

We adopt $T_\star=35000\K$ to characterize an ionizing spectrum
similar to a star of spectral type O8V, approximately representative
of the spectrum of stars powering typical \ion{H}{2} regions in 
star-forming galaxies.\footnote{E.g., $\theta^1$Ori~C (O7V) and
$\theta^1$Ori~D (O9.5V) in the Orion Nebula
\citep{ODell_2001}, with effective temperatures
$T_{\rm eff}=37000\,K$ and $32000\,K$.}
For $T_\star=35000\K$, the distribution (\ref{eq:phi_i}) 
has a mean photoelectron energy
\newtext{
\beq
\langle E_{\rm pe}\rangle = \sum_{i=1}^N \phi_i E_i \approx 4.22\eV
~.
\eeq
For $T_\star=1.0\xtimes10^5\K$, $\langle E_{\rm pe}\rangle = 15.83\eV$.
}

\subsection{\label{sec:free-free}
            Free-Free Radiation}

As a simple approximation, we assume that an electron of energy $E_i$
scattering off singly-charged ions with density $n_e$
radiates with power per unit frequency
\beqa
p_\nu(E_i)&=& A_0\,h\, n_e
\left(\frac{\eV}{E_i}\right)^{1/2} \hspace*{2.0cm} h\nu < E_i
\\
&=& 0 \hspace*{4.8cm} h\nu > E_j
~,
\eeqa
where $A_0$ is a constant.
With this assumption, a thermal distribution would have
\beq
\frac{4\pi j_\nu}{n_e} 
= \frac{2}{\sqrt{\pi}} A_0\,h\,n_e 
\left(\frac{\eV}{\kB T}\right)^{1/2} e^{-h\nu/\kB T}
~,~~
\eeq
and total radiated power per electron
\beqa
\frac{\Lambda}{n_e} &=& \frac{2}{\sqrt{\pi}} A_0 n_e
\left(\eV ~ \kB T\right)^{1/2}
\\
&=& 1.91\times10^{-25}\left(\frac{n_e}{\cm^{-3}}\right)
\left(\frac{T}{10^4\K}\right)^{1/2}\erg\s^{-1}
~,
\eeqa
for $A_0=1.108\times10^{-13}\cm^3\s^{-1}$.
This corresponds to
assuming a constant Gaunt factor $g_{\rm ff}=1.34$,
which is the frequency-averaged value near $T=10^4\K$.
[see, e.g., Eq.\ (10.10) in \citet{Draine_2011a}].

The probability per unit time of an electron of energy $E_i$
making a transition to bin $j<i$ is related to the radiated power by
\beq
(E_i-E_j) B_{ji} = p_\nu(E_i) \frac{\delta E}{h}
~.
\eeq
Thus, the probability per time of a transition
resulting from emission of a photon is
\beqa
B_{ji} &=& A_0 n_e\left(\frac{\eV}{E_i}\right)^{1/2}
\frac{\delta E}{(E_i-E_j)}
\\
&=&\frac{1.108\times10^{-13}\s^{-1}}{(E_i/\eV)^{1/2}}
\left(\frac{n_e}{\cm^{-3}}\right)
\frac{1}{(i-j)}
\hspace*{2.0cm}{\rm for~} i > j
~.
\eeqa

\subsection{\label{sec:collisional excitation}
            Collisional Excitation of Ions}

Electrons can undergo inelastic scattering with electron kinetic energy
going into electronic excitation of abundant ions such as
\ion{N}{2}, \ion{S}{2}, and \ion{O}{3}.
In Table \ref{tab:ion excitation} we list the
transitions included here, and the sources of inelastic cross sections.

\begin{table}[ht]
\begin{center}
\caption{\label{tab:ion excitation}
         Collisional Excitation Channels}

{\footnotesize
\begin{tabular}{c c c c c c}
\hline
X &        $\ell$ & $u$ & $(E_u-E_\ell)/hc$ & $(E_u-E_\ell)$ & reference\\
  &               &     &  ($\cm^{-1}$) & (eV) & \\
\hline
\ion{N}{2} & $^3$P$_0$             & $^3$P$_1$             & 48.7  & 0.00604 
           & \citet{Tayal_2011} \\
\ion{C}{2} & $^2$P$_{1/2}^{\rm o}$ & $^2$P$_{3/2}^{\rm o}$ & 63.4  & 0.00786 
           & \citet{Liang+Badnell+Zhao_2012} \\
\ion{O}{3} & $^3$P$_0$             & $^3$P$_1$             & 113.2 & 0.0140 
           & \citet{Tayal+Zatsarinny_2017}\\
\ion{N}{2} & $^3$P$_0$             & $^3$P$_2$             & 130.8 & 0.0162 
           & \citet{Tayal_2011} \\
\ion{N}{3} & $^2$P$_{1/2}^{\rm o}$ & $^2$P$_{3/2}^{\rm o}$ & 179.4 & 0.0222 
           & \citet{Liang+Badnell+Zhao_2012} \\
\ion{S}{3} & $^3$P$_0$             & $^3$P$_1$             & 298.7 & 0.0370 
           & \citet{Hudson+Ramsbottom+Scott_2012} \\
\ion{O}{3} & $^3$P$_0$             & $^3$P$_2$             & 306.2 & 0.0380 
           & \citet{Tayal+Zatsarinny_2017} \\
\ion{Ne}{3}& $^3$P$_2$             & $^3$P$_1$             & 642.9 & 0.0797 
           & \citet{McLaughlin+Bell_2000} \\
\ion{Ne}{2}& $^2$P$_{3/2}^{\rm o}$ & $^2$P$_{1/2}^{\rm o}$ & 780.4 & 0.0968 
           & \citet{Griffin+Mitnik+Badnell_2001} \\
\ion{S}{3} & $^3$P$_0$             & $^3$P$_2$             & 833.1 & 0.103 
           & \citet{Hudson+Ramsbottom+Scott_2012} \\
\ion{Ne}{3}& $^3$P$_2$             & $^3$P$_0$             & 920.6 & 0.114 
           & \citet{McLaughlin+Bell_2000} \\
\ion{S}{3} & $^3$P$_0$             & $^1$D$_2$             & 11323 & 1.404 
           & \citet{Hudson+Ramsbottom+Scott_2012} \\
\ion{N}{2} & $^3$P$_0$             & $^1$D$_2$             & 15316 & 1.899 
           & \citet{Tayal_2011} \\
\ion{O}{3} & $^3$P$_0$             & $^1$D$_2$             & 20273 & 2.513 
           & \citet{Tayal+Zatsarinny_2017} \\
\ion{Ne}{3}& $^3$P$_2$             & $^1$D$_2$             & 25841 & 3.204 
           & \citet{McLaughlin+Bell_2000} \\
\ion{O}{2} & $^4$S$_{3/2}^{\rm o}$ & $^2$D$_{5/2}^{\rm o}$ & 26811 & 3.324 
           & \citet{Tayal_2007} \\
\ion{O}{2} & $^4$S$_{3/2}^{\rm o}$ & $^2$D$_{3/2}^{\rm o}$ & 26831 & 3.327 
           & \citet{Tayal_2007} \\
\ion{S}{3} & $^3$P$_0$             & $^1$S$_0$             & 27161 & 3.367 
           & \citet{Hudson+Ramsbottom+Scott_2012} \\
\ion{N}{2} & $^3$P$_0$             & $^1$S$_0$             & 32689 & 4.053 
           & \citet{Tayal_2011} \\
\ion{O}{2} & $^4$S$_{3/2}^{\rm o}$ & $^2$P$_{3/2}^{\rm o}$ & 40468 & 5.017 
           & \citet{Tayal_2007} \\
\ion{O}{2} & $^4$S$_{3/2}^{\rm o}$ & $^2$P$_{1/2}^{\rm o}$ & 40470 & 5.017 
           & \citet{Tayal_2007} \\
\ion{C}{2} & $^2$P$_{1/2}^{\rm o}$ & $^4$P$_{1/2}$         & 43003 & 5.332 
           & \citet{Liang+Badnell+Zhao_2012} \\
\ion{C}{2} & $^2$P$_{1/2}^{\rm o}$ & $^4$P$_{3/2}$         & 43025 & 5.334 
           & \citet{Liang+Badnell+Zhao_2012} \\
\ion{C}{2} & $^2$P$_{1/2}^{\rm o}$ & $^4$P$_{5/2}$         & 43054 & 5.334 
           & \citet{Liang+Badnell+Zhao_2012} \\
\ion{O}{3} & $^3$P$_0$             & $^1$S$_0$             & 43186 & 5.354 
           & \citet{Tayal+Zatsarinny_2017} \\
\ion{C}{3} & $^1$S$_0$             & $^3$P$_0^{\rm o}$     & 52367 & 6.492 
           & \citet{Aggarwal+Keenan_2015} \\
\ion{C}{3} & $^1$S$_0$             & $^3$P$_1^{\rm o}$     & 52391 & 6.495 
           & \citet{Aggarwal+Keenan_2015} \\
\ion{C}{3} & $^1$S$_0$             & $^3$P$_2^{\rm o}$     & 52447 & 6.502 
           & \citet{Aggarwal+Keenan_2015} \\
\newtext{\ion{Ne}{3}}
           & \newtext{$^3$P$_2$}
           & \newtext{$^1$S$_0$}            
           & \newtext{55753} 
           & \newtext{6.913}
           & \newtext{\citet{McLaughlin+Bell_2000}} \\
\ion{N}{3} & $^2$P$_{1/2}^{\rm o}$  & $^4$P$_{1/2}$        & 57187 & 7.090
           & \citet{Liang+Badnell+Zhao_2012} \\
\ion{N}{3} & $^2$P$_{1/2}^{\rm o}$  & $^4$P$_{3/2}$        & 57247 & 7.098
           & \citet{Liang+Badnell+Zhao_2012} \\
\ion{N}{3} & $^2$P$_{1/2}^{\rm o}$  & $^4$P$_{5/2}$        & 57328 & 7.108
           & \citet{Liang+Badnell+Zhao_2012} \\
\hline
\end{tabular}
}
\end{center}
\end{table}
\begin{table}[t]
\begin{center}
\caption{\label{tab:parameters}
         Test Cases}
\begin{tabular}{c c c}
\hline
input parameter & H\,II region & Planetary Nebula \\
\hline
$T_\star$ (K) & $3.5\xtimes10^4$ & $1.0\xtimes10^5$\\
$n_e\,(\cm^{-3})$ & 1 & 1\\
$n_e/\nH$ & 1.1 & 1.1\\
$n({\rm C\,III})/\nH$ & $0.5\times 2.95\xtimes10^{-4}$ & $1.0\times 2.95\xtimes10^{-4}$\\
$n({\rm N\,II)}/\nH$  & $0.5\times 7.41\xtimes10^{-5}$ & $0$                           \\
$n({\rm N\,III})/\nH$ & $0.5\times 7.41\xtimes10^{-5}$ & $1.0\times 7.41\xtimes10^{-5}$\\
$n({\rm O\,II})/\nH$  & $0.4\times 5.37\xtimes10^{-4}$ & $0$\\
$n({\rm O\,III})/\nH$ & $0.4\times 5.37\xtimes10^{-4}$ & $1.0\times 5.37\xtimes10^{-4}$\\
$n({\rm Ne\,II})/\nH$ & $0.5\times 9.33\xtimes10^{-5}$ & $0.5\times 9.33\xtimes10^{-5}$\\
$n({\rm Ne\,III})/\nH$& $0.5\times 9.33\xtimes10^{-5}$ & $0.5\times 9.33\xtimes10^{-5}$\\
$n({\rm S\,III})/\nH$ & $1.0\times 1.45\xtimes10^{-5}$ & $1.0\times 1.45\xtimes10^{-5}$\\
$E_{\rm max}$ (eV)    & 25                             & 75 \\
$\langle E_{\rm pe}\rangle$ (eV)& $4.22$               & 15.83   \\
\hline
resulting $T_e$ (K)   &  6869                          & 11129 \\
\hline
\end{tabular}
\end{center}
\end{table}
\begin{figure}[ht]
\begin{center}
\includegraphics[angle=0,width=8.0cm,
                 clip=true,trim=0.5cm 5.0cm 0.5cm 2.5cm]
{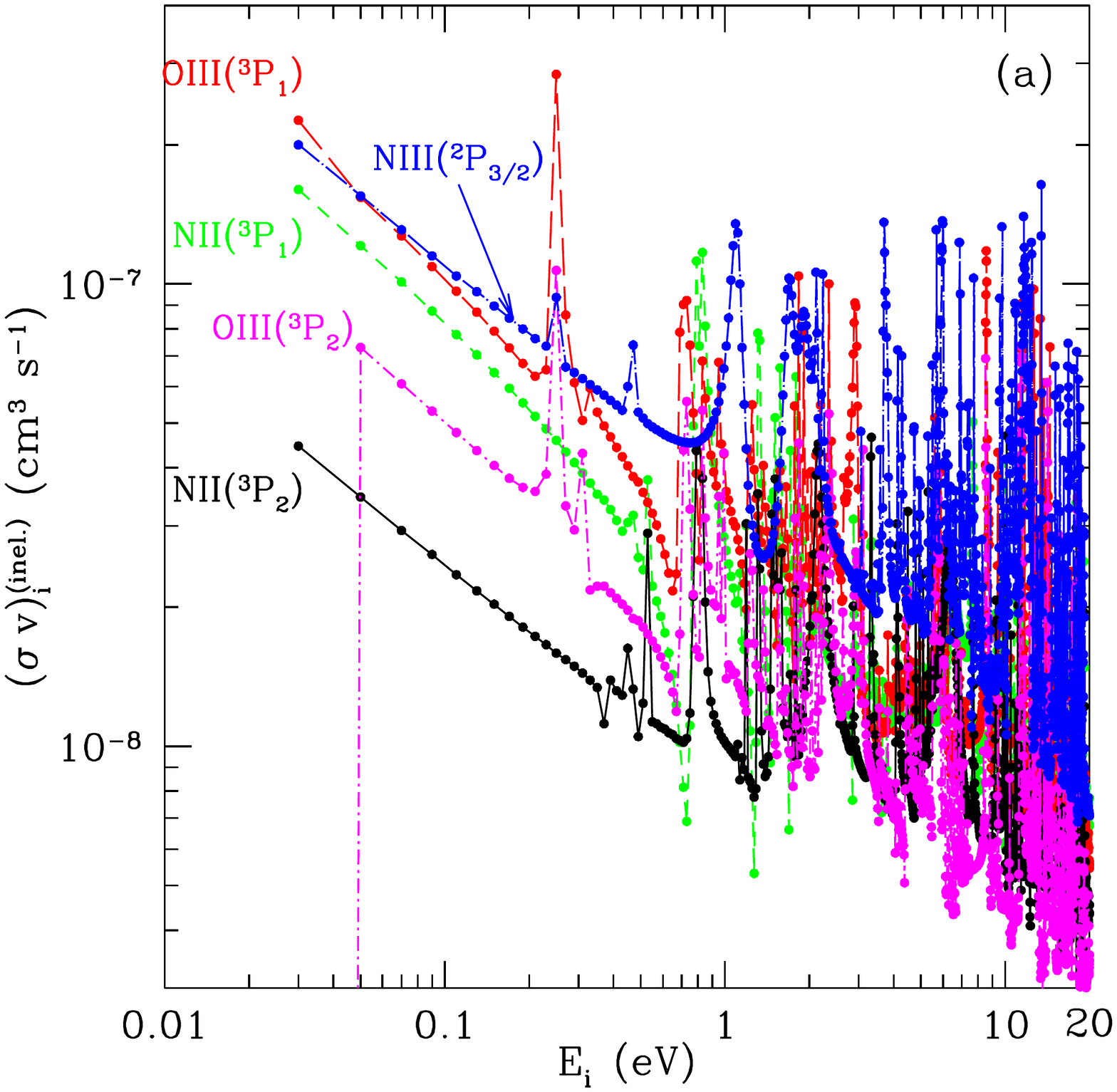}
\includegraphics[angle=0,width=8.0cm,
                 clip=true,trim=0.5cm 5.0cm 0.5cm 2.5cm]
{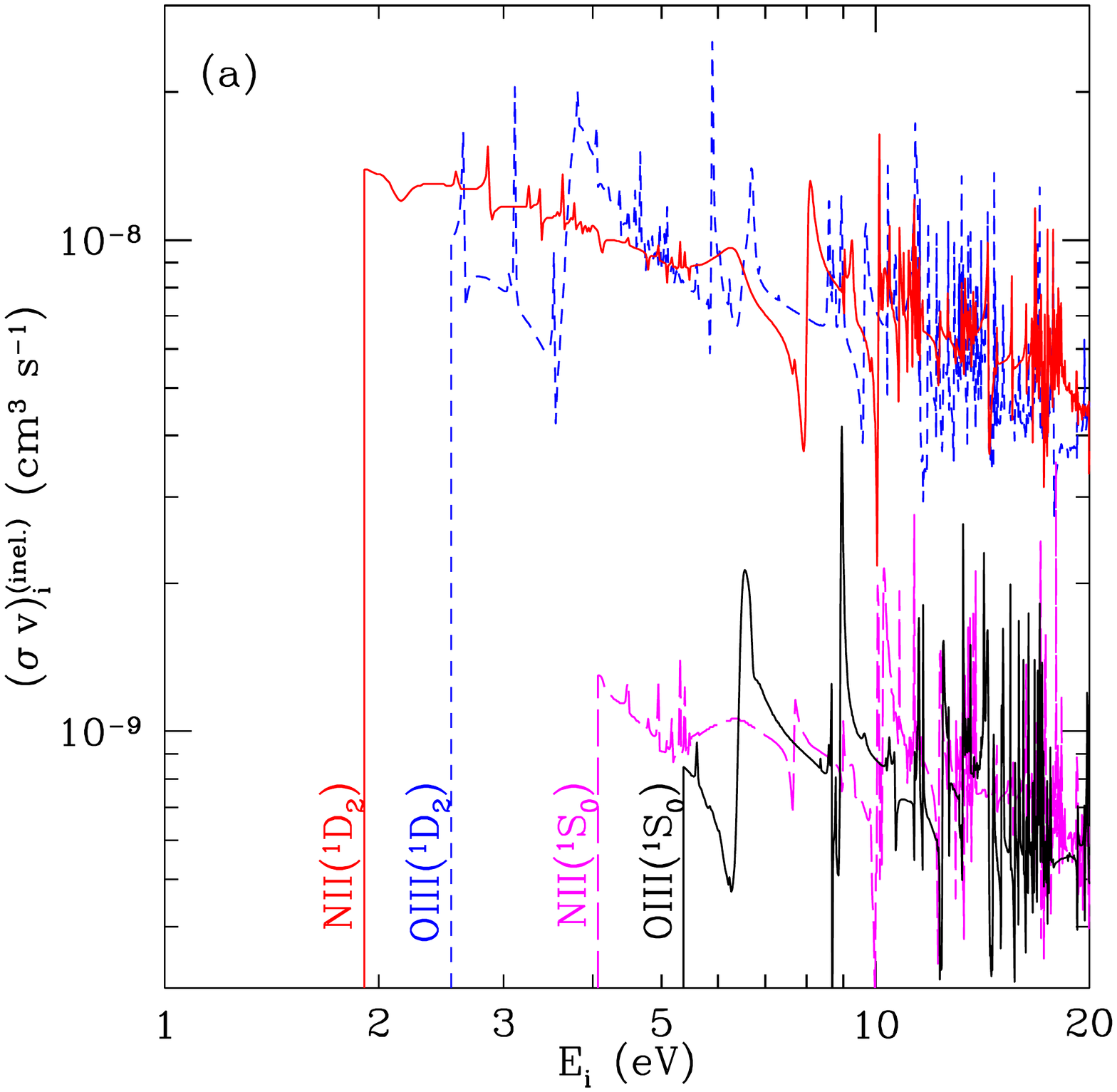}
\caption{\label{fig:inelastic scattering,fine structure} \footnotesize
         (a) Rate coefficients for fine structure excitation
         from the ground state 
         \newtext{$^3$P$_0$ to the excited fine structure levels of
         \ion{N}{2} and \ion{O}{3}}
         as a function of electron energy $E_i$.
         Dots show the discrete energies $E_j$ used in the calculations
         with $E_{\rm max}=25\eV$ and $N=1250$.  The lowest energy
         $E_1=0.010\eV$ is not shown, because energy loss is suppressed
         for electrons in this bin.
         \newtext{(b) Rate coefficients for electronic excitation of
         \ion{N}{2} and \ion{O}{3} from the ground state
         $^3$P$_0$.
         Cross sections are from \citet{Tayal_2011} and
         \citet{Tayal+Zatsarinny_2017}.}
         }
\end{center}
\vspace*{-0.5cm}
\end{figure}
\begin{figure}[ht]
\begin{center}
\includegraphics[angle=0,width=8.0cm,
                 clip=true,trim=0.5cm 5.0cm 0.5cm 2.5cm]
{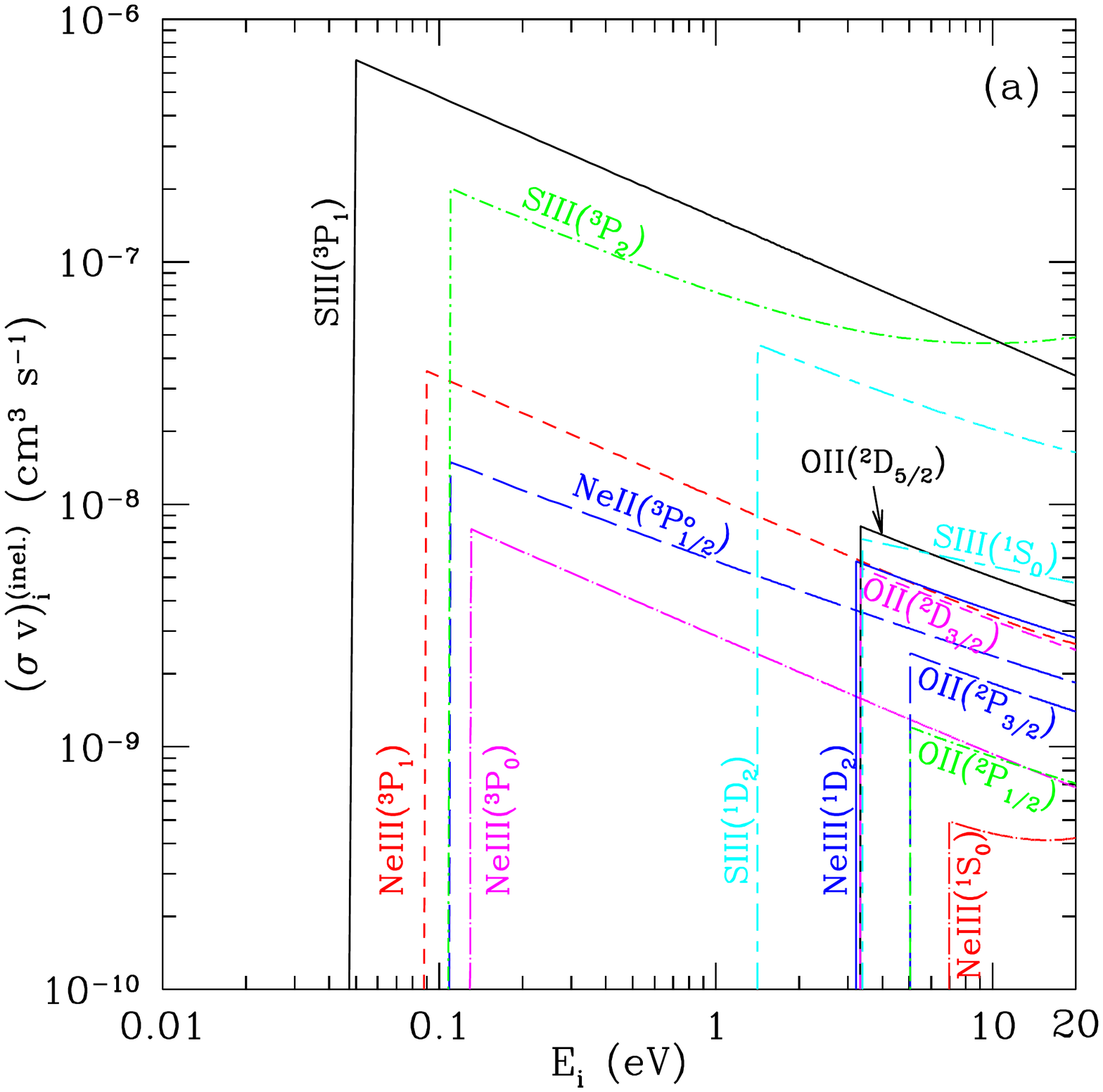}
\includegraphics[angle=0,width=8.0cm,
                 clip=true,trim=0.5cm 5.0cm 0.5cm 2.5cm]
{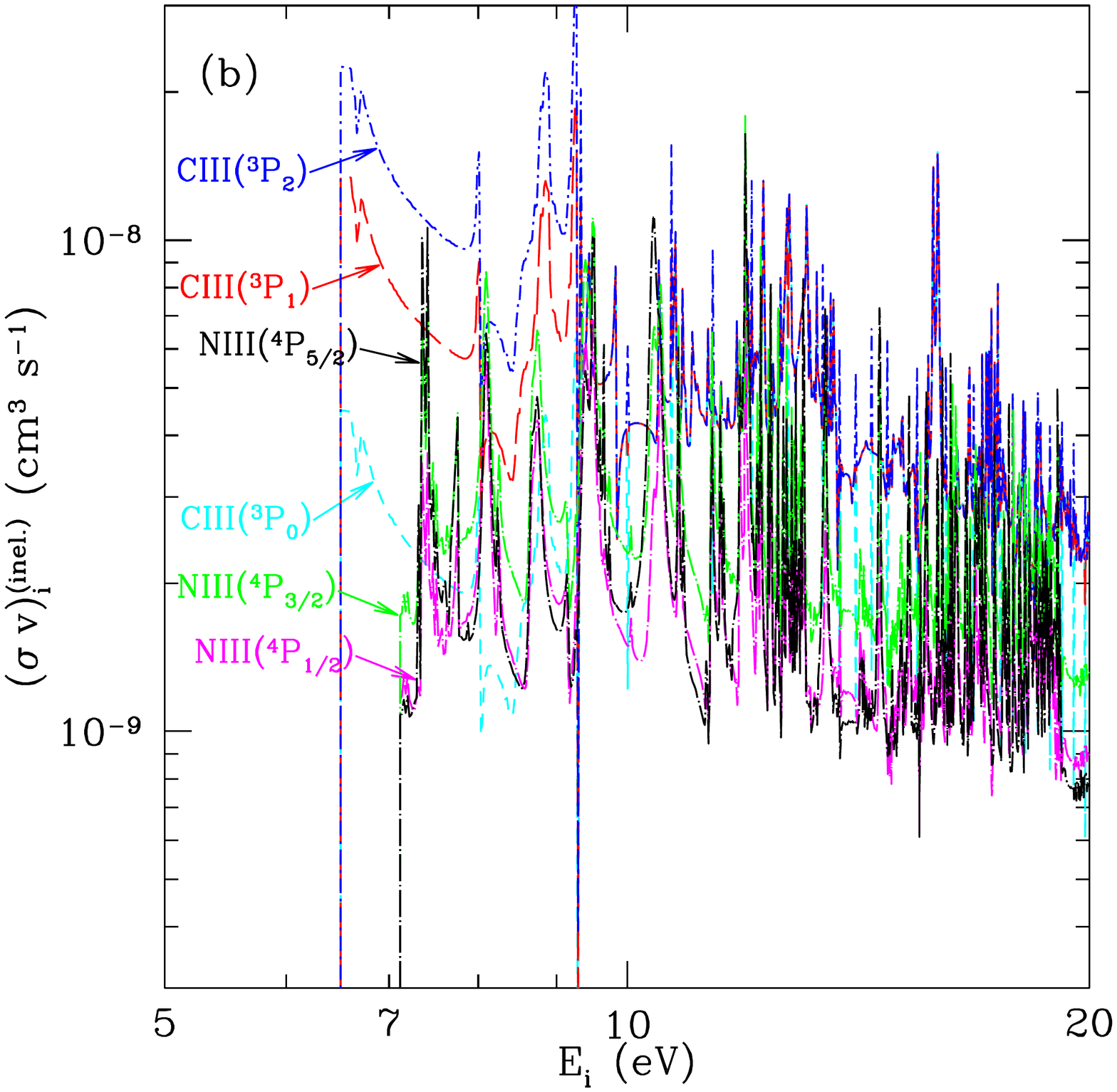}
\caption{\label{fig:inelastic scattering,electronic} \footnotesize
         Rate coefficients for excitation of excited electronic states
         as a function of electron energy $E_i$ (see text).
         \newtext{(a) Analytic fits to rates for
         \ion{O}{2} \citep{Tayal_2007},
         \ion{Ne}{2} \citep{Griffin+Mitnik+Badnell_2001},
         \ion{Ne}{3} \citep{McLaughlin+Bell_2000}, and
         \ion{S}{3} \citep{Hudson+Ramsbottom+Scott_2012}.
         (b) Rates for electronic excitation of
         \ion{C}{3} \citep{Aggarwal+Keenan_2015} and
         \ion{N}{3} \citep{Liang+Badnell+Zhao_2012}.}
         }
\end{center}
\vspace*{-0.5cm}
\end{figure}

Consider excitation of an ion $X$ from level $\ell$ to level $u$, requiring
excitation energy $\Delta E_{u\ell}$.
The cross section for excitation is expressed in terms of a dimensionless
energy-dependent collision strength $\Omega(E)$:
\beq
\sigma(E_i) = \pi a_0^2 \frac{\IH}{E_i}
\frac{\Omega_{\ell\rightarrow u}(E_i)}{g_\ell}
~~~{\rm for}~ E_i > \Delta E_{u\ell}
~
\eeq
where $g_\ell$ is the degeneracy of the lower level $\ell$.
The rate coefficient for inelastic scattering of an electron
of energy $E_i > \Delta E_{u\ell}$ is
\beq
(\sigma v)_i^{(\rm inel.)} 
= \pi a_0^2 \, \frac{\IH}{E_i} \, 
\frac{\Omega_{\ell\rightarrow u}(E_i)}{g_\ell} 
\left(\frac{2 E_i}{m_e}\right)^{1/2}
~.
\eeq
The collision strengths $\Omega_{\ell\rightarrow u}(E)$ were obtained from
the sources listed in Table \ref{tab:ion excitation}.
The resulting rate coefficients are shown in Figures
\ref{fig:inelastic scattering,fine structure} and
\ref{fig:inelastic scattering,electronic}.
We used the energy-dependent cross sections when available, giving
rate coefficients with sharp structure at resonances (e.g., resonances
in rates for excitation of the $^3$P$_1$ and $^3$P$_2$ levels of
\ion{O}{3} in 
Figure \ref{fig:inelastic scattering,fine structure}a).
When energy-dependent cross sections were not available, we used simple
fits that reproduced the calculated temperature-dependence of the
thermal rate coefficients 
(e.g., excitation of \ion{S}{3} $^3$P$_1$ and $^3$P$_2$ 
in 
Figure \ref{fig:inelastic scattering,electronic}a).

We assume that the density is low enough that excited states created by
collisional excitation decay by spontaneous emission of a photon before
collisional deexcitation can occur.
Thus, we neglect superelastic processes where excited states
are depopulated by collisions, and assume that every ion is
in the ground state before a scattering encounter.
This maximizes the collisional cooling rate relative to the photoionization
heating rate, with both then having the same scaling with density $n_e$, so that
the equilibrium temperature is independent of $n_e$.

Our treatment allows for fine-structure transitions 
which may be smaller than our energy bin width $\delta E$.
Let $k \equiv {\rm nint}(\Delta E_{u\ell}/\delta E)$, where ${\rm nint}(x)$
is the nearest integer.
If $k=0$ we take
\beq
C_{i-1,i} =  n_X \frac{\Delta E_{u\ell}}{\delta E} 
\times (\sigma v)_i^{\rm (inel.)}
\hspace*{1.5cm}{\rm for}~ i>1
~.~~
\eeq
If $k\geq 1$ we take
\beqa
C_{1,k+1} &=& n_X \frac{\Delta E_{u\ell}}{k\delta E} 
\times (\sigma v)_i^{\rm (inel.)}
\\
C_{i-k-1,i} &=& n_X f\times (\sigma v)_i^{\rm (inel.)} ~~~~~~~~~~{\rm if}~ i>k+1
\\
C_{i-k,i} &=& n_X (1-f)\times (\sigma v)_i^{\rm (inel.)} ~~~{\rm if}~ i>k+1
~,~
\eeqa
where $f\equiv (\Delta E_{u\ell}/\delta E) - k$, and
$n_X$ is the number density of the ion $X$.

These $C_{ji}$ give the correct rate of energy loss for electrons
with energy $E_i$.  For electrons in the lowest energy bin $i=1$ we
neglect inelastic energy loss.  So long as
$P_1\ll 1$ this underestimation of collisional cooling is small.
The final $C_{ij}$ used in our calculations are obtained by summing
over all of the cooling channels, using the ionic abundances $n_X/\nH$ 
listed in Table \ref{tab:parameters}.
Relative to the protosolar abundances of
\citet{Asplund+Grevesse+Sauval+Scott_2009}
we take C to be 50\% depleted, O to be 20\% depleted,
and N, Ne, and S to be undepleted.
We assume the gas phase C and S to be doubly-ionized,
and N, O, and Ne to be 50\% singly-ionized, and
50\% doubly-ionized.  Our objective is only to have a reasonable representation
of cooling processes and overall cooling rate, 
not to reproduce detailed ionization conditions at
any particular location.

\section{\label{sec:steady-state}
Steady-State Solutions \newtext{for \ion{H}{2} Region Conditions}}

\subsection{Numerical Solutions with Suppression of Elastic Scattering}

To examine the efficacy of electron-electron scattering,
we consider a modified problem where all elastic scattering processes
are suppressed by a factor $\gamma$.
Thus we seek $P_i$ satisfying the modified equations
\beqa \nonumber 
0 &=&
\left(\sum_{j=1}^N R_j P_j\right) \phi_i
-
R_i P_i
+
\sum_{j=i+1}^N \left(B_{ij}+C_{ij}\right)P_j
-
\sum_{i=1}^{j-1} \left(B_{ji}+C_{ji}\right)P_i
\\ \nonumber
&&+\gamma
\Bigg[
\sum_{j=1}^{i-1} ~ \sum_{k=i-j+1}^N   A_{i-j,j,k} P_j P_k 
+\sum_{k=i+1}^N \sum_{j=1}^{N-(k-i)} ~ A_{k-i,j,k} P_j P_k 
\
\\ \label{eq:modified dPdt}
&& ~~~
-\sum_{j=1}^{N-i} ~ \sum_{k=j+1}^N A_{j,i,k} P_i P_k 
-\sum_{j=1}^{i-1} ~ \sum_{k=1}^{N-j} A_{j,k,i} P_k P_i
\Bigg] 
~.
\eeqa
For $\gamma=1$ we recover the original physical problem, but by 
considering smaller values of $\gamma$ we can illustrate the
importance of elastic scattering compared to energy-changing processes.

\begin{figure}[ht]
\begin{center}
\includegraphics[angle=0,width=10.0cm,
                 clip=true,trim=0.5cm 5.0cm 0.5cm 2.5cm]
{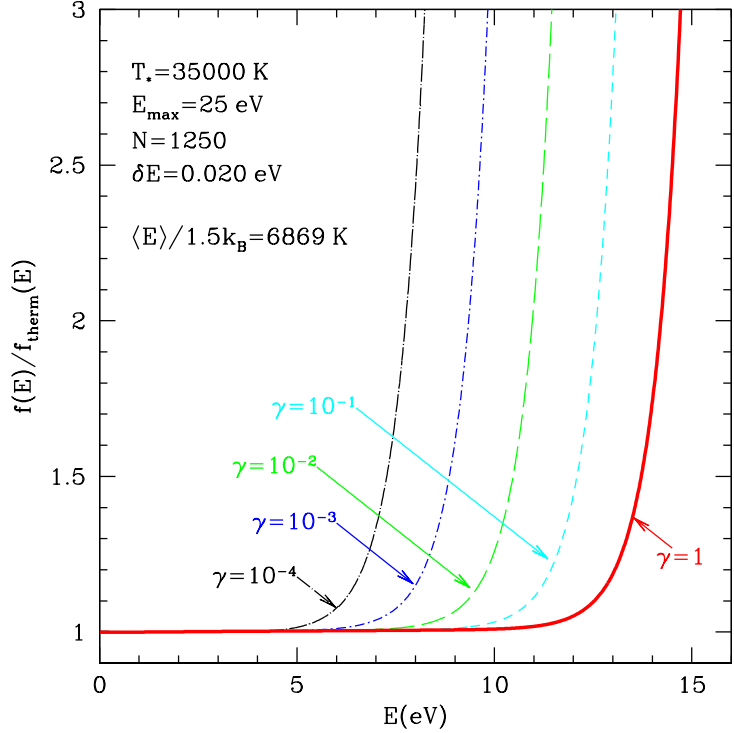}
\caption{\label{fig:steady-state}\footnotesize
Solid curve: Steady-state electron distribution function relative to
Maxwellian for the parameters in Table \ref{tab:parameters}.  
The energy distribution is accurately described by
a Maxwellian for $E\ltsim 13\eV$.
Broken curves: distribution functions if elastic scattering
is artificially suppressed by a
factor $\gamma$.
We see that even if elastic scattering is suppressed by a
factor $10^{-4}$, it is still able to maintain a Maxwellian
distribution below $7\eV$.
}
\end{center}
\vspace*{-0.5cm}
\end{figure}
\begin{figure}[ht]
\begin{center}
\includegraphics[angle=0,width=10.0cm,
                 clip=true,trim=0.5cm 5.0cm 0.5cm 2.5cm]
{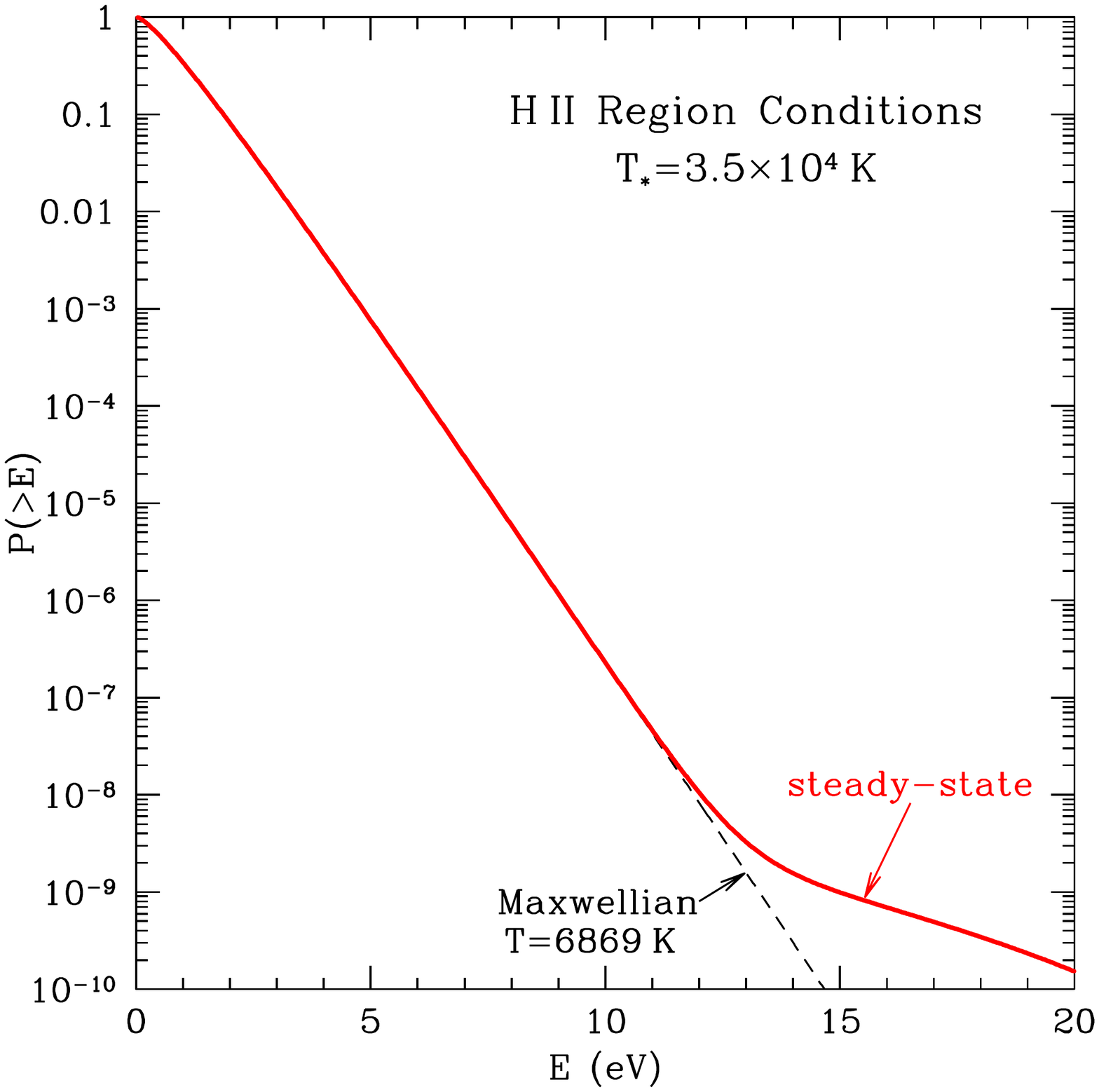}
\caption{\label{fig:cde}\footnotesize
         Fraction of electrons with energy $>E$, as a function
         of $E$, for the steady-state solution.
         Only $\sim$$2\xtimes10^{-9}$ of the electrons are
         in the nonthermal ``tail'' of the distribution at
         $E \gtsim 13\eV$.
         }
\end{center}
\end{figure}

Starting from an initial guess for the $P_j$,
the steady-state solution $P_j$ is found by iteration using
the Fortran implementation of the Levenberg-Marquard algorithm from
the minpack library
\citep{Garbow+Hillstrom+More_1980}.
Because elastic scattering is very fast compared to the energy-changing
processes (photoionization, recombination, free-free emission, and
collisional excitation) the Levenberg-Marquard algorithm, even using
64 bit arithmetic, only reaches an approximate steady-state.
Therefore, after the Levenberg-Marquard algorithm has reached the
limit of its numerical accuracy, we take the resulting $P_i$ as
a starting point and evolve forward in time.
We have obtained accurate solutions to Eq.\ (\ref{eq:modified dPdt})
for various values of $\gamma$.
Figure \ref{fig:steady-state} shows results for 5 values of $\gamma$,
ranging from $\gamma=10^{-4}$ to $\gamma=1$.

The resulting statistical steady-state is in thermal equilibrium, with
heating balancing cooling.
The mean electron energy is $\langle E \rangle \equiv \sum P_j E_j$.
We define the ``temperature'' to be
\beq
T \equiv \frac{\langle E \rangle}{1.5\kB}
~.
\eeq
In Figure \ref{fig:steady-state} we show 
the electron energy distributions $P_j$ relative to a Maxwellian
distribution
\beq \label{eq:MB dist}
\left(\frac{dP}{dE}\right)_{\rm MB}
= \frac{2}{\sqrt{\pi}} 
\frac{\sqrt{E}}{(\kB T)^{3/2}}e^{-E/\kB T}
~.
\eeq
Even with elastic scattering suppressed by a factor $\gamma=10^{-4}$,
$P_j$ accurately follows the Maxwell-Boltzmann distribution up to
$E=5\eV$, although there is a high energy tail that exceeds the
Maxwell-Boltzmann distribution.  As elastic scattering becomes
stronger, it is able to maintain a thermal distribution of electron energies
up to higher energies.  For the physical case of $\gamma=1$, the
electrons are thermally-distributed up to $\sim$$13\eV$, or
$E/\kB T_{\rm eff} \approx 20$.

Figure \ref{fig:cde} shows the fraction of the electrons above
energy $E$, as a function of $E$.
The steady-state distribution is very close to a Maxwellian up to
$\sim13\eV$.
Only a very small
fraction  --  approximately $2\xtimes10^{-9}$ -- of the
electrons are in the high energy tail maintained by
steady injection of energetic photoelectrons.

\subsection{$\kappa$-Distributions}

The $\kappa$-distribution is \citep{Vasyliunas_1968}
\beq
\left(\frac{dP}{dE}\right)_\kappa =
\frac{2}{\sqrt{\pi}}
\frac{1}{(\kappa-\frac{3}{2})^{3/2}}
\frac{\Gamma(\kappa+1)}{\Gamma(\kappa-\frac{1}{2})}
\frac{\sqrt{E}}{(\kB T_U)^{3/2}}
\left(1 + \frac{E}{(\kappa-\frac{3}{2})\kB T_U}\right)^{-(\kappa+1)}
~,
\eeq
where $T_U \equiv (2/3)\langle E\rangle/\kB$ is the effective temperature,
and $\kappa$ is a dimensionless parameter.
In the limit $\kappa\rightarrow\infty$ the $\kappa$-distribution
converges to the Maxwell-Boltzmann distribution (\ref{eq:MB dist}),
but for $3/2 < \kappa < \infty$
the $\kappa$-distribution has relatively more electrons at high
energies than the Maxwell-Boltzmann distribution with the same
mean kinetic energy.

\begin{figure}[ht]
\begin{center}
\includegraphics[angle=0,width=10.0cm,
                 clip=true,trim=0.5cm 5.0cm 0.5cm 2.5cm]
{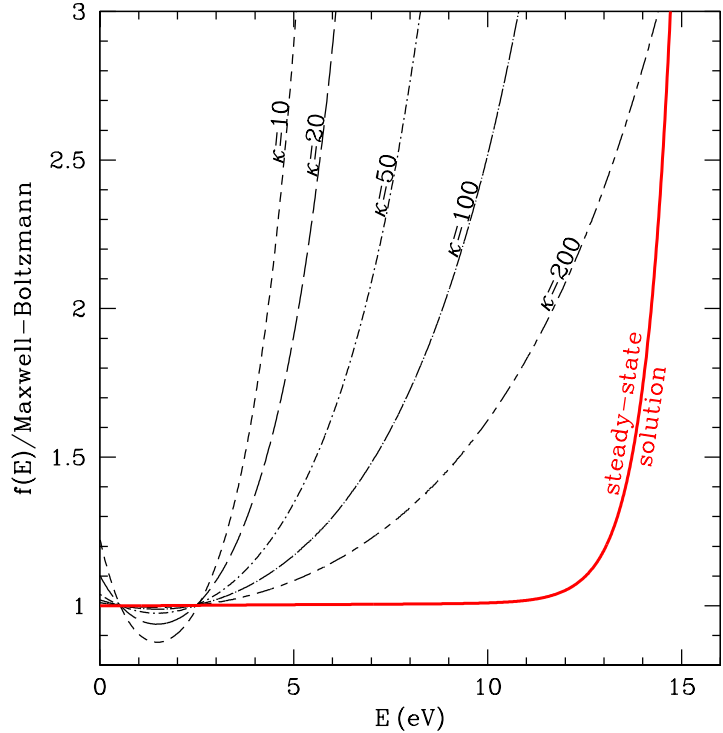}
\caption{\label{fig:kappa} \footnotesize
         Steady-state electron energy distribution function (solid curve).
         Also shown are $\kappa$-distributions for $T_U=6869\K$ and
         $\kappa=50$, 100, and 200.
         Even for $\kappa=200$ the $\kappa$-distribution seriously
         overestimates the electron energy distribution between $5$ and $14\eV$.
         The Maxwellian distribution is an excellent approximation up to
         $13\eV$.
         }
\end{center}
\end{figure}

\citet{Vasyliunas_1968} used $\kappa$-distributions to describe
electron energy distributions in the Earth's magnetosphere
measured by the OGO-1 and OGO-3 satellites, and since then 
$\kappa$-distributions have often been used to describe
``space plasmas''.
\citet{Livadiotis+McComas_2011} discuss the theoretical
basis for the $\kappa$-distribution
as a non-equilibrium stationary state.

\citet{Nicholls+Dopita+Sutherland_2012} argue that 
the observed line ratios in
\ion{H}{2} regions are the result of $\kappa$-distributed electrons,
with values of the $\kappa$ parameter in the
range $10\ltsim\kappa\ltsim 20$.
For planetary nebulae they found $\kappa\gtsim10$ from comparison
of temperatures determined from [\ion{O}{3}]4364/[\ion{O}{3}]5009
with temperatures determined from the Balmer and Paschen breaks
in the hydrogen recombination radiation.

In Figure \ref{fig:kappa} we compare our calculated steady-state
solution for the electron energy distribution with $\kappa$-distributions.
All cases have the same mean electron kinetic energy
$\langle E\rangle = 0.8879\eV$ (i.e., $T_U=6869\K$).
As in Figure \ref{fig:steady-state}, 
we show the energy distribution function divided by
a Maxwellian distribution.
Even for $\kappa=200$, the $\kappa$-distribution substantially
overestimates the fraction of electrons with energies in the
5--12$\eV$ range.
Up to 12$\eV$, the electron energy distribution is very accurately
described by a Maxwellian, with factor-of-two departures present only
above 14$\eV$.  

\section{\label{sec:relaxation}
         Time-Dependent Solutions: Relaxation to the Stationary State}

To illustrate the speed of relaxation to a near-Maxwellian distribution,
we calculate the evolution of the electron energy distribution, starting
from an initial distribution that is highly non-Maxwellian.
We use the same physics: photoionization=recombination, recombination
cross sections as in Section \ref{sec:recombination}, and cooling by
free-free emission and collisional excitation as in
Sections \ref{sec:free-free} and \ref{sec:collisional excitation}.

Our objective here is only to show the speed of relaxation, so the
initial conditions are arbitrary.  While we include both cooling by
free-free emission and line excitation, and photoionization balancing
recombination, on the short time scales on which
elastic scattering is able to thermalize the electron distribution, the
energy of the plasma is nearly constant.  For convenience, we choose
an initial distribution with the same mean energy per particle as in 
the steady state solution 
with heating balancing cooling
(corresponding to $T=6869\K$), but we start with
90\% of the electrons in a cold Maxwellian distribution with 50\% of the total
energy, and 10\% of the electrons in a Maxwellian distribution with
the remaining 50\% of the energy.  
Thus the cooler Maxwellian has a temperature
$T=3816\K$, and the hotter Maxwellian has a temperature
$T=34345\K$.

\begin{figure}[ht]
\begin{center}
\includegraphics[angle=0,width=8.0cm,
                 clip=true,trim=0.5cm 5.0cm 0.5cm 2.5cm]
{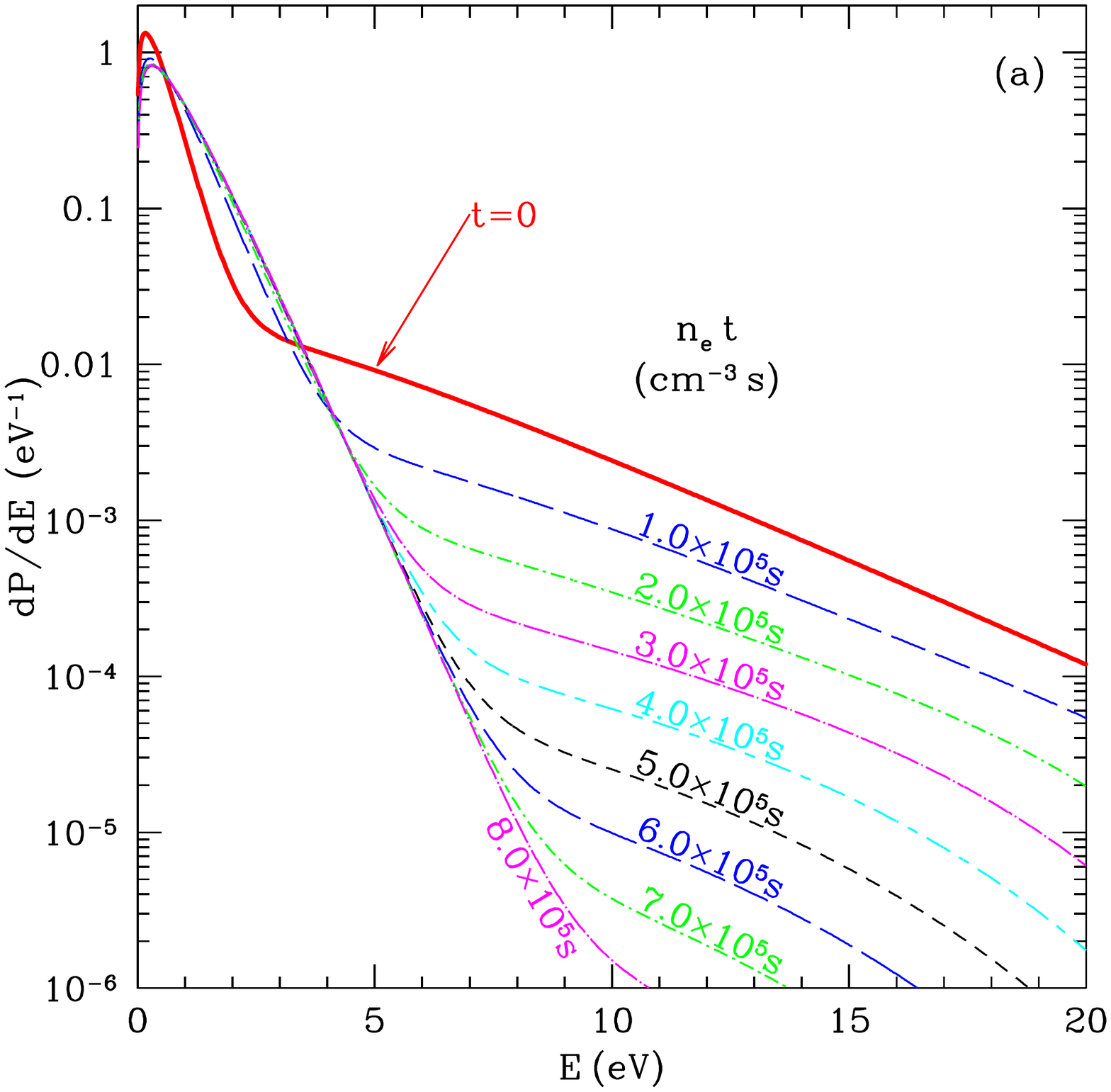}
\includegraphics[angle=0,width=8.0cm,
                 clip=true,trim=0.5cm 5.0cm 0.5cm 2.5cm]
{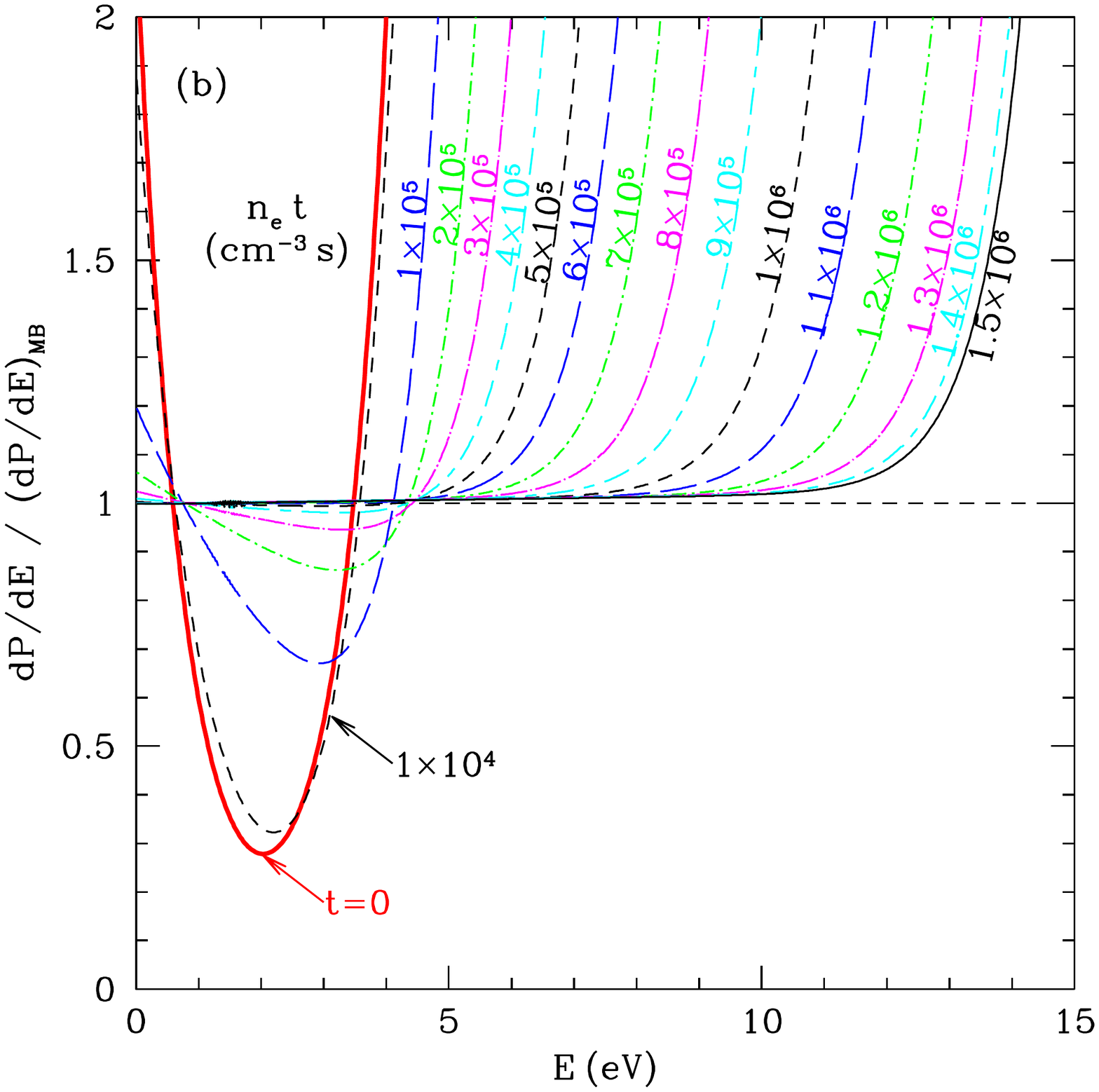}
\caption{\label{fig:relax}\footnotesize
         Relaxation toward a thermal distribution.
         (a) Electron energy distribution $dP/dE$ at selected times.
         (b) Electron energy distribution relative to a Maxwellian,
         at selected times.
         For $E<13\eV$, the distribution has relaxed to a Maxwellian
         after only $1.5\xtimes10^6\s$ for $n_e=1\cm^{-3}$.
         }
\end{center}
\end{figure}

Figure \ref{fig:relax} shows evolution of the electron energy
distribution, starting from the initial 
distribution at $t=0$.
The high energy tail relaxes on a time scale 
\beq \label{eq:tau_relax, HII}
\tau_{\rm relax}\approx 10^5
\left(\frac{\cm^{-3}}{n_e}\right)\sec
~.
\eeq
At the densities of the Orion nebula ($n_e\approx 3000\cm^{-3}$), 
$\tau_{\rm relax} \approx 30\s$ -- relaxation is extremely fast!

For this example, which begins with an extreme excess at high
energies, the abundance of $\sim10\eV$ electrons drops by 4 orders of
magnitude in $\sim$$10^6(\cm^{-3}/n_e)\s$, and for $E\ltsim 13\eV$
is accurately described by a Maxwellian after only 
$1.5\xtimes10^6 (\cm^{-3}/n_e)\s$.
A time $\sim$$10\tau_{\rm relax}$ is required
to reduce the extreme high energy tail by a factor $\sim$$10^4$.

\newtext{

\section{\label{sec:PN} 
         Planetary Nebulae}

The nuclei of planetary nebulae have photospheric temperatures
as high as $\sim2\xtimes10^5\K$.
Consequently, the plasma receives much more electron kinetic
energy per photoionization than in the \ion{H}{2} regions photoionized
by O stars.
The increased heating per photoionization results in higher
gas temperatures.

Electron temperatures in the range 8000--20000\,K
have been estimated from the nebular and auroral lines of \ion{O}{3}
\citep{Zhang+Liu+Wesson+etal_2004}.
The line ratios in planetary nebulae often appear to be inconsistent
with a single electron temperature, and the discrepancies have
sometimes been attributed to $\kappa$-distributed electron energies
\citep{Zhang+Zhang+Liu_2016}.

\begin{figure}[ht]
\begin{center}
\includegraphics[angle=0,width=8.0cm,
                 clip=true,trim=0.5cm 5.0cm 0.5cm 2.5cm]
{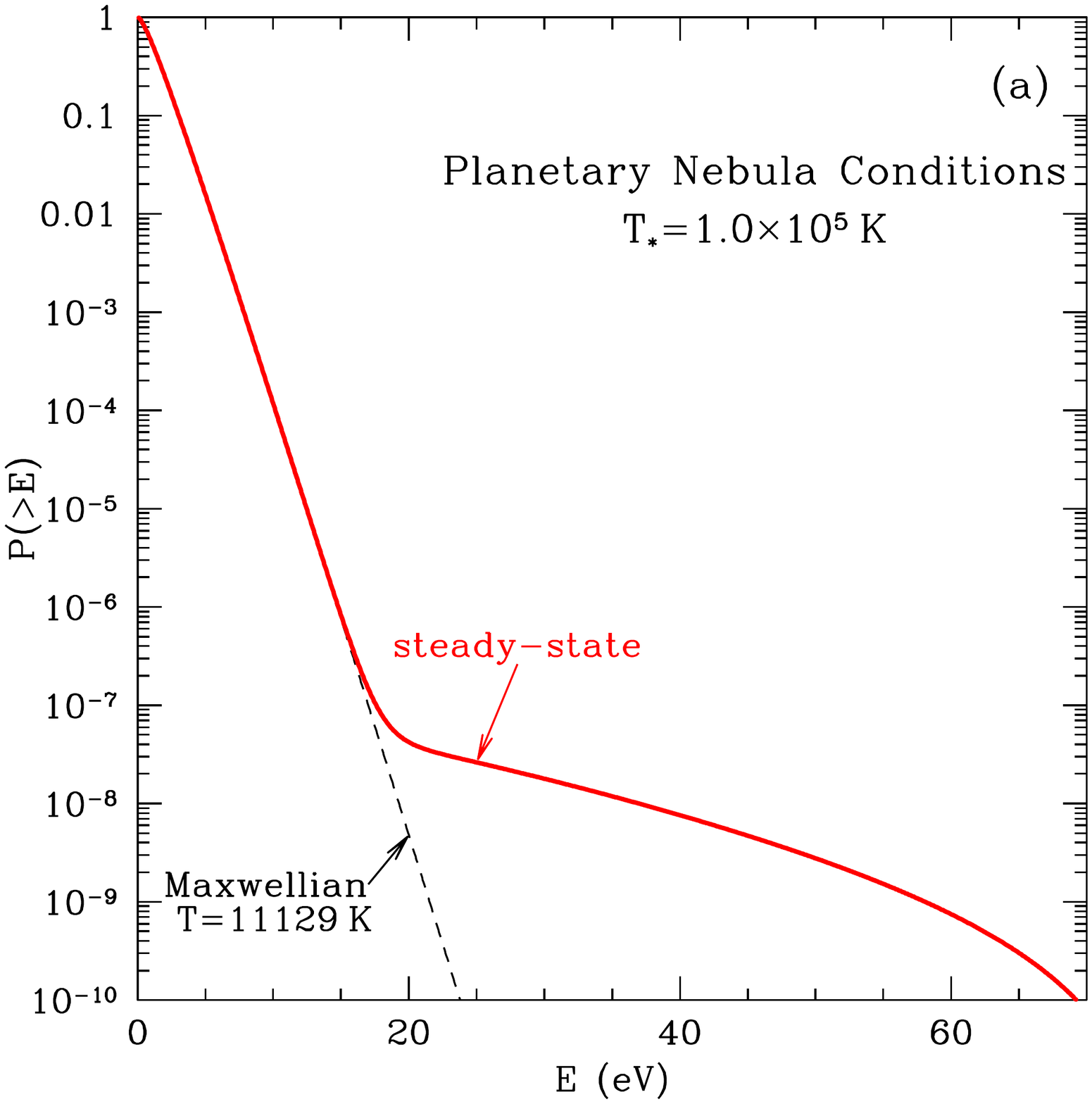}
\includegraphics[angle=0,width=8.0cm,
                 clip=true,trim=0.5cm 5.0cm 0.5cm 2.5cm]
{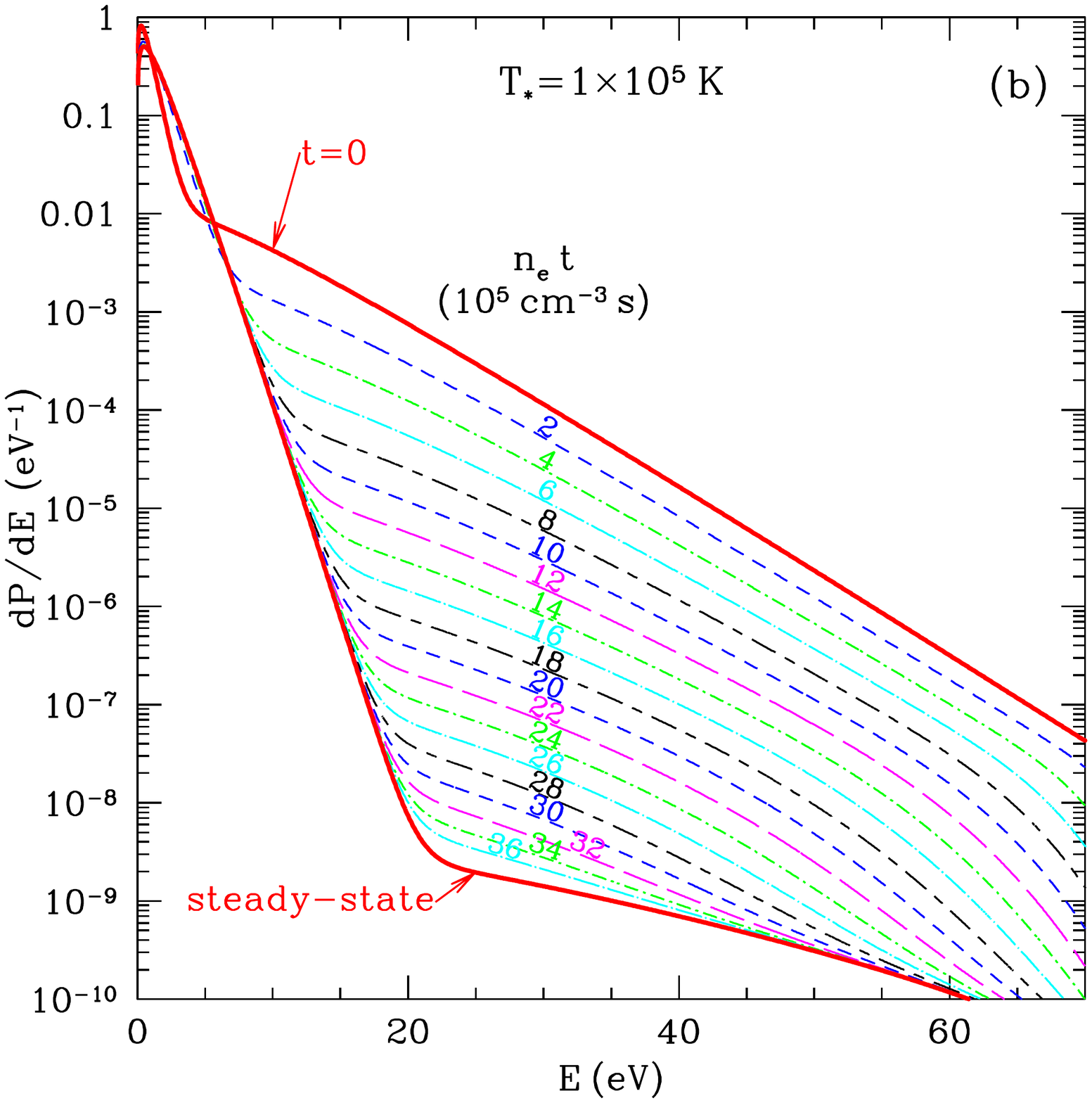}
\caption{\footnotesize\label{fig:PN_cde}
\newtext{
(a) Steady-state fraction of electrons with energy $>E$ as a function of $E$,
for the average spectrum of photoelectrons
for a stellar temperature $T_\star=10^5\K$.
Up to $E=16\eV$ the distribution is accurately described by
a Maxwellian, but $\sim5\xtimes10^{-8}$ of the electrons
are in a high-energy tail beginning at about $\sim19\eV$.
(b) Time-dependent solutions, starting from a dual Maxwellian at $t=0$
with 10\% of the electrons containing 90\% of the energy.
The distribution relaxes toward the steady-state solution on
a timescale $\sim2\xtimes10^5 (\cm^{-3}/n_e)\s$.
}
}
\end{center}
\end{figure}

Here we calculate the steady-state electron temperature in
a plasma heated by the hard spectra present in planetary nebulae.
As an example, we consider a stellar temperature
$T_\star=1.0\xtimes10^5\K$, as for the central star
of the Helix Nebula NGC\,7291
\citep{Napiwotzki_1999}.
We use the average spectrum of photoelectron energies from
Eq.\ (\ref{eq:phi_i}); for $T_\star=10^5\K$, the mean
energy of photoelectrons from H ionization is 
$\langle E_{\rm pe}\rangle=15.8\eV$.

Figure \ref{fig:PN_cde}a
shows the calculated cumulative steady-state electron energy distribution,
compared to a Maxwellian with $T=11129\K$.
Nearly all of the electrons, up to
$E\approx 16\eV$, are distributed following a
Maxwellian distribution with $T=11129\K$, but $\sim3\xtimes10^{-8}$ of the
electrons are in a high energy tail extending to higher energies.
The electrons in this tail were recently injected with 
energies $E>20\eV$, and are
in the process of slowing down to join the thermal distribution.

Figure \ref{fig:PN_cde}b
shows time-dependent relaxation toward the steady-state solution,
for an initial distribution at $t=0$ with the same mean energy per particle
as in the steady-state solution, but with 90\% of the energy contained
in 10\% of the particles (assumed to be in a Maxwellian distribution).
Relaxation toward the steady-state solution takes place with a
relaxation time 
\beq
\tau_{\rm relax}\approx 2\xtimes10^{5}\left(\frac{\cm^{-3}}{n_e}\right)\s
~.
\eeq
The relaxation time is longer than for \ion{H}{2} region conditions
(Eq.\ \ref{eq:tau_relax, HII}) because the mean energy per particle
is larger by a factor $(11129/6869)=1.62$, and the energy equipartition
time for Coulomb scattering scales as $E^{1.5}$.
For this illustrative example, a time $\sim$$15\tau_{\rm relax}$
is needed to become close to the steady state for $E\gtsim 25\eV$,
because the assumed initial conditions need to decay by a factor
$\sim$$10^6$ to reach the steady state.
Nevertheless, the time
$\sim$$3\xtimes10^6(\cm^{-3}/n_e)\s$ is short compared to other time scales.
}

\section{Summary}

Local relaxation to a near-Maxwellian energy distribution is
very rapid for the conditions in \ion{H}{2} regions and planetary nebulae.
There is no basis for using
$\kappa$-distributions to describe the electrons in \ion{H}{2}
regions \newtext{or planetary nebulae.}  

Given the speed of thermal relaxation 
($\tau_{\rm relax}\approx30\sec$ for Orion Nebula
conditions),
if observed line ratios and recombination spectra
are found to be inconsistent with a single-temperature Maxwellian, 
this must be the result of the observed spectra summing
emission from regions with different temperatures.
Observations of real \ion{H}{2} regions often include emission from
an ionization front bounding the ionized gas, where photoionization
can substantially exceed recombination, allowing heating rates
to exceed the steady-state value.
In addition to the ionization front at the outer boundary of the
\ion{H}{2} region, there may also be ionization fronts around
dense neutral globules within the \ion{H}{2} region that are undergoing
``photoevaporation''.
\newtext{Such ionization fronts may locally have electron temperatures
well above the temperatures in the bulk of the photoionized gas,
where photoioinization is limited by the rate of
radiative recombination, thus limiting the photoelectric heating.}

The shorter time scales for photoionization in ionization fronts
propagating into neutral gas will allow the electron population to have
nonthermal ``tails'' that are larger than in the steady-state,
but given that only $\sim$$2\xtimes10^{-9}$ of the electrons are in the
nonthermal tail for steady-state photoionization (see Figure \ref{fig:cde})
\newtext{in \ion{H}{2} regions (or $\sim$$5\xtimes10^{-8}$ in
planetary nebulae)},
it seems unlikely that the population of the nonthermal tail will
be large enough to 
\newtext{significantly} affect spectra even in propagating photoionization
fronts.

Thus observations of \ion{H}{2} regions and planetary nebulae
should be interpreted using a mixture of local temperatures
(and ionization conditions)
as originally proposed by \citet{Peimbert_1967}.
The electrons are locally well-approximated by Maxwellian distributions -- 
$\kappa$-distributions do not apply.

\acknowledgments
\newtext{We thank Gary Ferland for helpful suggestions 
that led to improvement of this paper.}
We 
are grateful to
Kanti Aggarwal, 
Guiyun Liang, 
Brendan McLaughlin, 
Swaraj Tayal, and 
Oleg Zatsarinny
for providing inelastic cross section data.

This work made use of the references provided in the CHIANTI
atomic database \citep{Dere+Landi+Mason+etal_1997,DelZanna+Dere+Young+etal_2015}.
This research was supported in part by NSF grant AST-1408723,
and in part by NSF Research Fellowship DGE~1656466 to CDK.

\bibliography{/u/draine/work/bib/btdrefs}
\begin{appendix}
\section{\label{app:ee scattering}
         Electron-Electron Scattering}

The electron-electron scattering problem is straightforward
kinematics plus Rutherford scattering.  Because the algebra is
somewhat involved, we collect the results here.

Let $\bv_j$ and $\bv_k$ be the electron velocities before
scattering, with $\bv_j\cdot\bv_k=v_jv_k\cos\Theta$.
After scattering the velocities are $\bv_j^\prime$ and $\bv_k^\prime$.
The relative speed is
\beq
|\bv_j-\bv_k| = \left(\frac{2}{m}\right)^{1/2}
\left[E_j+E_k-2\sqrt{E_jE_k}\cos\Theta\right]^{1/2}
~.
\eeq
The energy gain by particle $j$ is
\beqa
\Delta E_j &=& \alpha \sqrt{E_jE_k}
\\
\alpha(\theta) &=& \beta(1-\cos\theta)+\gamma\sin\theta\
\\
\beta &\equiv&  \frac{E_k-E_j}{2\sqrt{E_jE_k}}
\\
\gamma &\equiv& \sin\Theta\cos\phi ~,
\eeqa
where $\theta$ is the scattering angle in the center-of-mass frame,
and $0\leq\phi\leq2\pi$ is the angle between the $\bv_j-\bv_k$ plane and
the $\bv_j^\prime-\bv_k^\prime$ plane.

\section{\label{app:A_ijk}
         $A_{ijk}$ for $i>1$ and $k>j$}

For an electron with initial energy $E_j$, scattering off electrons with
energy $E_k$,
the rate of energy transfer in scattering events with
$\Delta E \equiv E_j^\prime - E_j$ satisfying
\beq \label{eq:DeltaE condition}
\Delta E \in \left[\left(i-\frac{1}{2}\right),\left(i+\frac{1}{2}\right)\right]\delta E
\eeq
is
\beq
i\delta E \times n_k A_{ijk}
= n_k \int_0^\pi \!\frac{\sin\Theta d\Theta}{2} ~ |\bv_j-\bv_k| \int^{\prime\prime} 
\!\! d\Omega ~\Delta E \frac{d\sigma}{d\Omega}
~,
\eeq
where $n_k$ is the number density of electrons with energy $E_k$, and
$\int^{\prime\prime}$ is limited to $(\theta,\phi)$ such that
Eq.\ (\ref{eq:DeltaE condition}) is satisfied.
Let
\beq
Q_{\ell u}(E_j,E_k,\Theta,\phi) \equiv \int^\prime \!\!d\theta\sin\theta
\Delta E_j \frac{d\sigma(\theta)}{d\Omega}
~,
\eeq
where $\int^\prime$ is limited to $\theta$ values such that
Eq.\ (\ref{eq:DeltaE condition}) is satisfied, i.e.,
\beq
\alpha(\theta) \in 
\left[\left(i-\frac{1}{2}\right),\left(i+\frac{1}{2}\right)\right]
\frac{\delta E}{\sqrt{E_jE_k}} ~~~{\rm for}~i>1
~.
\eeq
Then
\beq
A_{ijk} = 
\frac{1}{2i\delta E} \int_0^\pi \!\sin\Theta d\Theta ~
|\bv_j-\bv_k|
\int^\prime \!\!d\phi ~
Q_{\ell u}(E_j,E_k,\Theta,\phi)
~.
\eeq
For Rutherford scattering, we have
\beq
Q_{\ell u}(E_j,E_k,\Theta,\phi)
=
\sqrt{E_jE_k} 
\left(\frac{e^2}{4 E_{\rm CM}}\right)^2
\int^\prime \!\! d\theta 
\frac{\sin\theta}{\sin^4(\theta/2)}
\left[
\beta(1-\cos\theta)+\gamma\sin\theta
\right]
~.
\eeq

\section{\label{app:A_1jk}
         $A_{ijk}$ for $i=1$ and $k>j$}
Because of the importance of weak, small-angle scattering,
evaluation of $A_{ijk}$ requires special treatment when
$i=1$.
Because we want all of the weak energy transfer events included,
we include all events with
\beq
\Delta E \in \left[0\,,\,\frac{3}{2}\right]\delta E ~,
\eeq
or
\beq
\alpha(\theta) \in \left[0 \,,\, \frac{3}{2}\right]\frac{\delta E}{\sqrt{E_jE_k}} ~~~{\rm for}~i=1
~.
\eeq

\section{\label{app:detailed balance}
            $A_{ijk}$ for $k\leq j$ from Detailed Balance}

The principle of microscopic reversibility requires that
\beq
g_j g_k A_{ijk} = g_{k-i}g_{j+i}A_{i,k-i,j+i} ~,
\eeq
where $g_j$ is the ``degeneracy'' of energy bin $j$.
Each of our ``bins'' includes a range of energies.  $g_j$ is
proportional to the number of quantum states within the energy interval,
with
\beq
g_j \propto \int^\prime v^2 dv ~,
\eeq
where $\int^\prime$ extends over kinetic energies in 
$[E_j-\delta E/2,E_j+\delta E/2]$.  Thus
\beqa
g_j&\propto& \int^\prime E^{1/2} dE = 
\frac{2}{3}\left[\left(E_j+\frac{\delta E}{2}\right)^{3/2}
-\left(E_j-\frac{\delta E}{2}\right)^{3/2}\right]
\\
&\propto& E_j^{1/2}\delta E 
\left[1-\frac{1}{96}\left(\frac{\delta E}{E_j}\right)^2 + 
O\left(\frac{\delta E}{E_j}\right)^4\right]
~.
\eeqa
Thus, to leading order in $\delta E/E_j$,
\beq \label{eq:reversibility}
A_{ijk} \approx \frac{E_{j+i}^{1/2}\,E_{k-i}^{1/2}}{E_j^{1/2}\,E_k^{1/2}}
\,A_{i,k-i,j+i}
=\sqrt{\frac{(j+i-\frac{1}{2})(k-i-\frac{1}{2})}
               {(j-\frac{1}{2})(k-\frac{1}{2})}}
\,A_{i,k-i,j+i}
~.
\eeq
Once we compute all the $A_{ijk}$ for $j<k$, we can obtain
the remaining $A_{ijk}$ using Eq.\ (\ref{eq:reversibility}). 

\end{appendix}
\end{document}